\documentclass[structabstract]{aa}

\usepackage{graphicx}

\usepackage{txfonts}

\begin{document}
   \title{Stochastic model of optical variability of BL Lac}

   \author{D. A. Blinov\inst{1,2}
        \and
         V. A. Hagen\textendash Thorn
          \inst{1,2}
          }
   \institute{St.-Petersburg State Univ., Universitetsky pr. 28, 198504 St.-Petersburg, Russia
         \and
             Isaac Newton Institute of Chile, St.-Petersburg Branch
             }

   \date{Received June 10, 2009; accepted June 20, 2009}

  \abstract
   {We use optical photometric and polarimetric data of BL Lacertae that cover a period of 22
years to study the variability of the source.}
   {The long-term observations are employed for establishing parameters of a stochastic
model consisting of the radiation from a steady polarized source and a number of 
variable components with different polarization parameters, proposed by the authors early.} 
   {We infer parameters of the model from the observations using numerical simulations based
on a Monte Carlo method, with values of each model parameter selected from a Gaussian distribution.
We determine the best set of model parameters by comparing model distributions to the observational ones using the $\chi^2$ criterion. }
   {We show that the observed photometric and polarimetric variability can be explained within a model
with a steady source of high polarization, $\sim$40\%, and with direction of polarization 
parallel to the parsec scale jet, along with 10$\pm$5 sources of variable polarization. }
   {}

\authorrunning{D.A. Blinov \& V.A. Hagen-Thorn}
   \maketitle
\makeatletter{\renewcommand*{\@makefnmark}{}
\footnotetext{{\it Send offprint requests to}: dmitriy.blinov@gmail.com}\makeatother}

\section{Introduction}
\object{BL Lacertae} is an active galactic nucleus (z=0.069) that is the prototype of a class of 
AGN, BL Lac's objects, which are characterized by strong dominance of non-thermal continuum,
extreme optical variability, and high optical polarization (\cite{AS80}). BL~Lac objects
belong also to the class of blazars with highly variable emission from radio
wavelengths to $\gamma$-rays and superluminal apparent velocities in radio jets.
Recent theoretical and observational results (e.g., \cite{VK04,McK06,MA08})
suggest that the jets are magnetically dominated, with a helical structure of the magnetic field  
close to the central engine and accretion disk where the jet formation and acceleration take place. 
However, at some distance from the central engine
the jets become kinetically dominated, with a turbulent magnetic field that is subject to ordering
by shocks. If the non-thermal optical emission in blazars originates close to the millimeter-wave VLBI
core, as suggested by \cite{LS00,GAB06,J07} then the optical polarization behavior 
should reflect the dynamics of the magnetic field in this region. 
 
On long-term timescales, many BL~Lac objects show a preferential direction of optical polarization
(\cite{HT80,HT02} - Paper 1) closely aligned with the direction of the parsec scale 
jet (\cite{J07}), as well as some correlations between total intensity and polarization
parameters (Paper~1). The data on OJ 287 (\cite{HT80}) and BL Lac (Paper~1) permit to suggest
a phenomenological model of variability for BL~ Lac object:
there is a constantly acting source of polarized radiation responsible for the preferential direction
of observed polarization and a number of sources with randomly distributed polarization
parameters and intensities that are superimposed on the steady source.

In this paper we use simultaneous photometric and polarimetric measurements of BL~Lac
obtained over 22 years (for details see Paper~1) and develop a method to derive the optimal number 
of the superimposed sources as well as the properties of the steady and superimposed sources
that give the best fit to the data. We apply the results to describe the
magnetic field structure in the optical emission region.

We recall that the relative Stokes parameters $p_x$ and $p_y$ are directly measured from polarimetric observations, and they are used to derive the degree of polarization $P=(p_x^2+p_y^2)^{1/2}$ and
position angle of polarization $\Theta_0=0.5\arctan(p_y/p_x)$. Photometric measurements
were performed independently and were corrected for contribution of the host galaxy.
The transformation from magnitudes {\it m} to intensities {\it I} was performed with the absolute calibration
of \cite{Mead90}. Polarimetric and photometric measurements obtained on the same
night were used to compute the absolute Stokes parameters $I, Q=p_x I,$ and $U=p_y I$.

\section{Stochastic Model}

Figures 1-6 ({\it left column}) show the observed dependences of 
polarization parameters on Stokes parameter $I$. Figure 7a presents the plot
between the degree of polarization and position angle of polarization, while
Figure 8a gives the dependence between the relative Stokes parameters.

The aim of this work is to reproduce all dependences obtained from the observations
with a model consisting of a steady source of polarized radiation with parameters 
$I^c, p_x^c, p_y^c$  and  \textit{n} variable sources with parameters 
$I^v_i, p^v_{xi}, p^v_{yi}$, where $i=1,\ldots,n$.
At a given date the observed optical emission is represented
by Stokes parameters $I_{mod}, Q_{mod}$, and $U_{mod}$ that are computed as the superposition
of the constantly acting and variable sources (see Appendix for the basic equations). 
The total number of dates is 451.

All parameters are assumed to be normally distributed around their mean value (mathematical
expectation M) with standard deviation $\sigma$. There are three constraints on the parameters of
the steady source: 1) its flux must be less than the minimum observed flux,
2) its polarization direction must yield the preferred direction of polarization, which is
$\Theta_0^{pref}=24^{\circ}$, and 3) its degree of polarization must be sufficiently high
to ensure a high observed polarization al low flux levels.

Determination of optimal model parameters was accomplished by a Monte
Carlo method. After multiple realizations of the model with different sets of parameters,
we accept a set of parameters that produces the best agreement with the observed
dependences as the most probable model. The initial criterion for rejecting 
obviously inadequate models and outlining possible sets of parameters was the visual similarity
between the modeled and observed correlations. 

The details of modelling were as follows: 1) the space of Stokes parameters
$\{I, Q, U\}$ was separated into \textit{K} non-intersecting regions, each containing about 10
points from the observed set of 451 points; 2) from possible sets of parameters
a realization of values $M(n), \sigma(n), M(I^v), \sigma(I^v), M(p_x^v), \sigma(p_x^v), M(p_y^v),
 \sigma(p_y^v), I^c, p_x^c, p_y^c$ was selected by chance (the value of $p_y^c$ was calculated from the
condition $\Theta_0^c = 24^{\circ}$, and it was assumed that $\sigma(p_y^c) = \sigma(p_x^c)$);
3) using this realization we found by random chance the values $n, I^v, p_x^v, p_y^v, I^c,
p_x^c, p_y^c$ for 451 cases; 4) using equations (\ref{absolute1}) - (\ref{lasteq}) for each of the
451 cases, we determined $I_{mod},Q_{mod},U_{mod}$; and 5) we compared the number of points
in each of \textit{K} regions derived in the model, $MOD_i$ ($i=1,\ldots,K$), with the number of observed 
points, $OBS_i$ ($i=1,\ldots,K$), by calculating the value of $\chi^2$:
\begin{equation}
 \chi^2 = \sum_{i=1}^K \frac{(MOD_i - OBS_i)^2}{OBS_i}.
\end{equation}

After repeating the third step 1000 times, we have found $\bar{\chi^2}$ for the realization chosen 
in the second step. Regarding $\bar{\chi^2}$ as a function of parameters of the second step,
we have determined the set of parameters that gives a global minimum of $\bar{\chi^2}$ by the genetic
algorithm of optimization (\cite{GenAlg}).
We used the {\it PyGene} library to implement the algorithm.
Beginning with the second step, the calculations were performed 200 times. The whole
algorithm was repeated for 3 different separations of \{$I, Q, U$\}-space. This gives 600 values of parameters. The mean values that give the most probable realization are as follows: $M(n)=10\pm1, \sigma(n)=5\pm1, M(I^v)=0.12\pm0.04,
 \sigma(I^v)=0.15\pm0.05, M(p_x^v)=2\pm1, \sigma(p_x^v)=16\pm3, M(p_y^v)=2\pm1, \sigma(p_y^v)=18\pm3,
 I^c=0.19\pm0.03, p_x^c=27\pm5, p_y^c=28\pm4$.

\section{Results and Conclusions}
We have constructed the dependences between the polarization parameters and intensity similar to the
observed dependences using such set of parameters:
$M(n)=10, \sigma(n)=5, M(I^v)=0.12, \sigma(I^v)=0.15, M(p_x^v)=2, \sigma(p_x^v)=16, M(p_y^v)=2,
 \sigma(p_y^v)=18, I^c=0.19, p_x^c=27, p_y^c=28$.
Figures 1b-8b give the modeled dependences obtained
for one realization of parameters taken at random from this set.
There is an excellent agreement between the observed and modelled
dependences. The similarity is excellent for other realizations of this set of parameters as well. 
Figures \ref{P_Iadd},\ref{P_Thadd},
as examples, give two dependences obtained for two more different realizations.
In addition, since the value of $\sigma(I^v)$ is higher than $M(I^v)$ in some cases 
we formally may obtain by chance a negative value of the intensity $I^v$; we exclude such cases.

The proposed stochastic model agrees very well with the phenomenological model suggested in Paper~1.
The model gives a high degree of polarization for the steady source, $P\sim40$\%,
with direction of polarization along the jet. This can be produced by a toroidal
component of a helical magnetic field in the magnetically dominated part of the jet. Such a scenario implies
that the optical emission most likely comes from the acceleration zone of the jet, as suggested by 
\cite{MA08}, who recently have observed in BL~Lac steady rotation of the position angle of optical 
polarization by about 240$^\circ$ over a five-day interval. The rotation can be explained within 
a helical field structure located in the acceleration zone $\sim$0.4~pc upstream
of the mm-wave VLBI core. However, the intensity of the steady source is only $\sim$50\% 
higher than the most probable intensity of a random variable source, while at a given moment we expect $\sim$10 such sources to operate simultaneously.
The most probable polarization of variable sources $\sim$3\%, which
implies that each variable source is an emission region with a significant number of cells with randomly
oriented magnetic field. This means that, in general, the magnetic field is dominated
by a turbulent component that masks the toroidal component. However, $\sigma p_x^v$ and $\sigma p_y^v$
are fairly high, which requires ordering of the magnetic field  in the emission regions
of the variable sources at some times. The latter can be connected with propagation of disturbances
along the jet and formation of shocks in the optical emission region. 

\renewcommand{\figurename}{Figs.}
\renewcommand{\thefigure}{1-7}
    \begin{figure*}[!ph]
    \centering
   \includegraphics[angle=270,width=17.1cm]{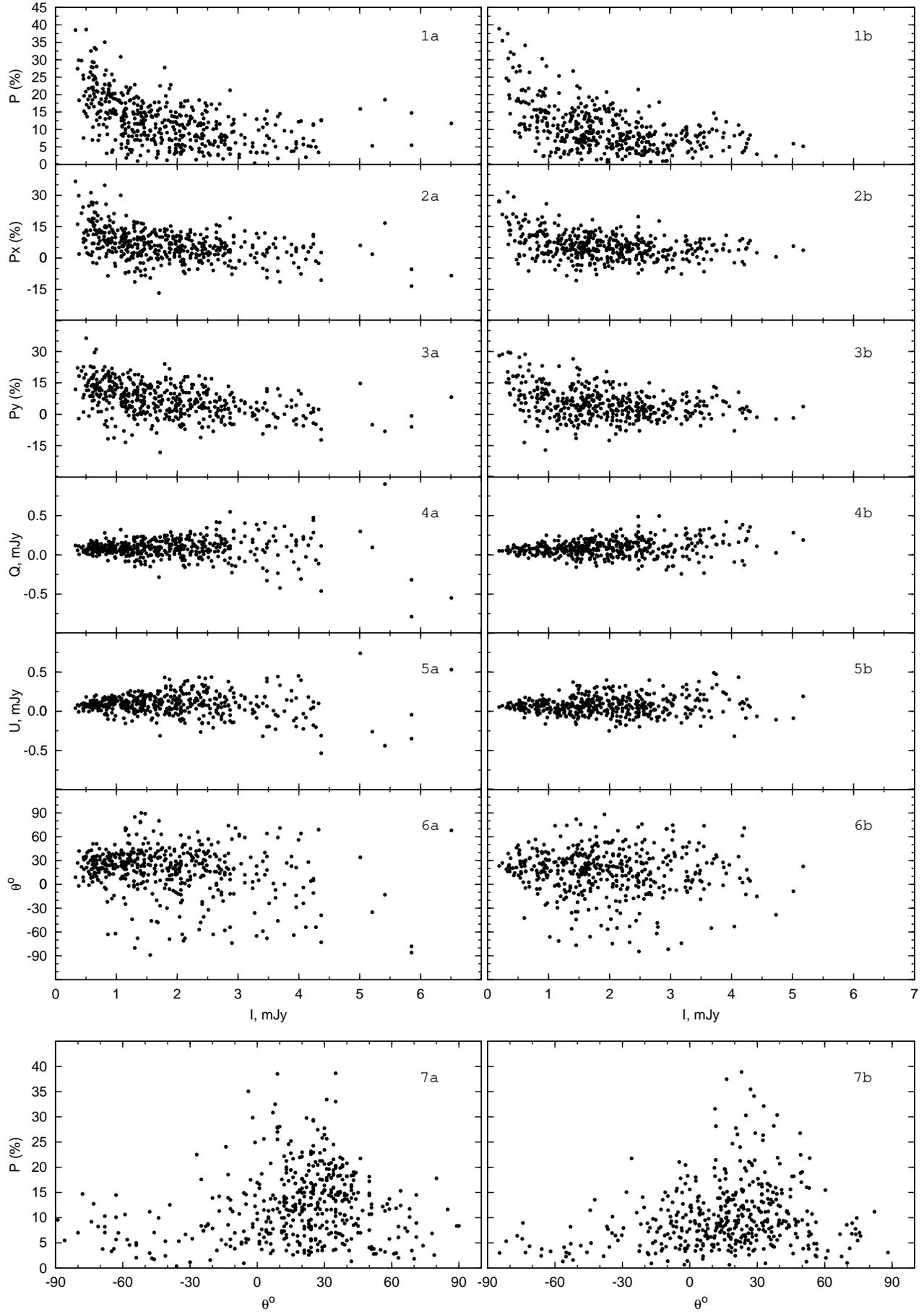}
    \caption{Observed dependences (left column) and modeling results (right column) between polarization
parameters and intensity.}
               \label{FigGam1}%
     \end{figure*}

\renewcommand{\figurename}{Fig.}
\renewcommand{\thefigure}{8}

\begin{figure*}[!t]
   \includegraphics[angle=270,width=16cm]{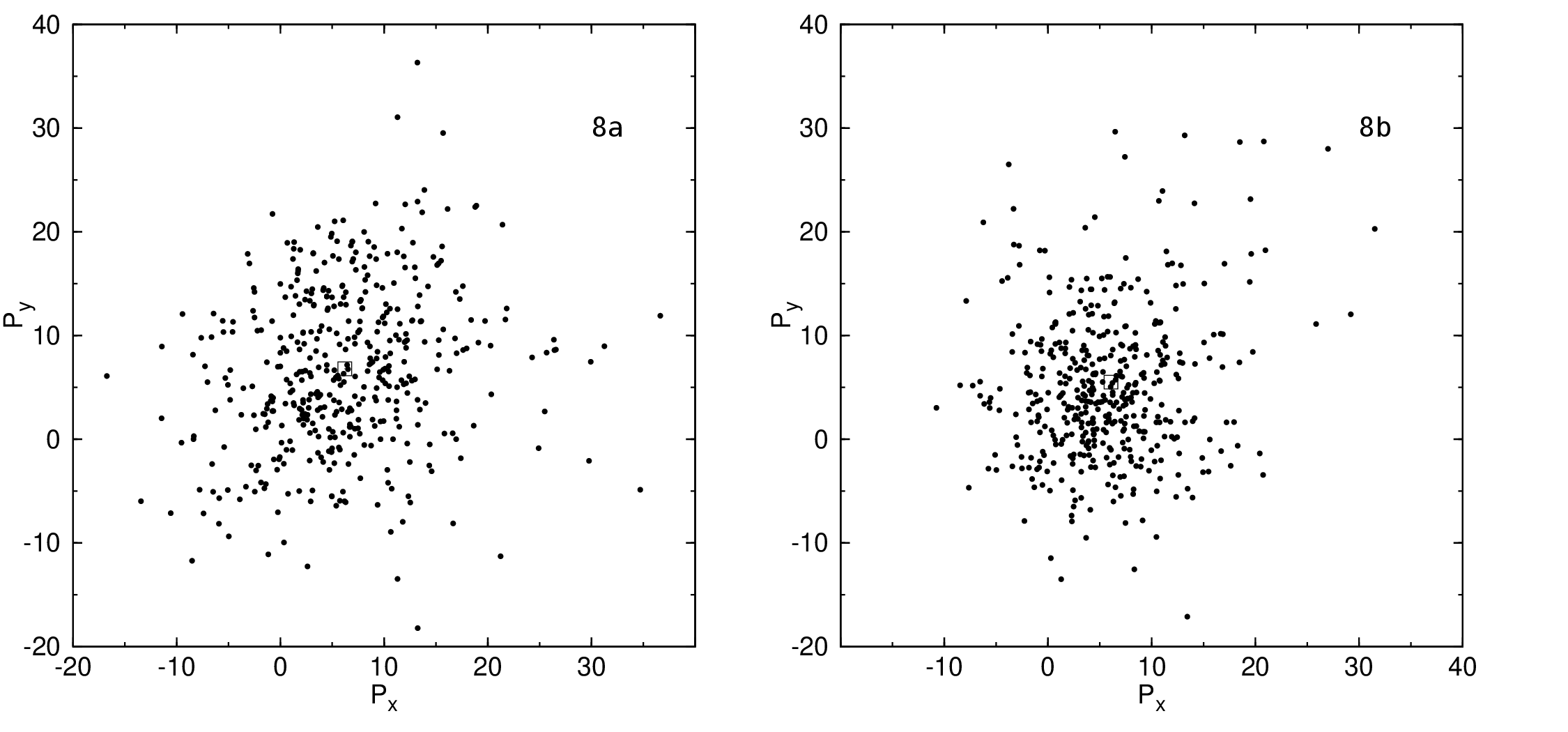}
\caption{ Relative Stokes parameters: observed (left) and modeled (right).
          Mean values are indicated by open squares.}
               \label{FigPxPy}%
     \end{figure*}

\setcounter{equation}{0}
\renewcommand{\theequation}{A\arabic{equation}}

\section{Appendix}
Because of additive nature of absolute Stokes parameters, the main equations
for obtaining the model values are as follows:
 \begin{equation}
 I_{mod} = I^c +  \sum_{i=1}^n I^v_i ,       \label{absolute1}                \\
 \end{equation}
 \begin{equation}
 \ Q_{mod} = p_x^c I^c +  \sum_{i=1}^n p^v_{xi} I^v_i, \\
 \end{equation}
 \begin{equation}
 \ U_{mod} = p_y^c I^c +  \sum_{i=1}^n p^v_{yi} I^v_i.
\label{absolute3}
 \end{equation}
From (\ref{absolute1}) - (\ref{absolute3}) for relative Stokes parameters we have
 \begin{equation}
    p_{x\,mod} =   \frac{p_x^c I^c +\sum_{i=1}^n p^v_{xi} I^v_i}{I^c + \sum_{i=1}^n I^v_i},  \\
 \end{equation}
 \begin{equation}
 \  p_{y\,mod} =   \frac{p_y^c I^c +\sum_{i=1}^n p^v_{yi} I^v_i}{I^c + \sum_{i=1}^n I^v_i},
 \end{equation}
and obviously,
 \begin{equation}
   p_{mod} =   \sqrt{p^2_{x\, mod}+p^2_{y\, mod}},  \\
 \end{equation}
 \begin{equation}
 \ \Theta_{0\,mod} =   \frac{1}{2} \arctan{\left(\frac{p_{y\, mod}}{p_{x\, mod}}\right)}.
\label{lasteq}
 \end{equation}

\begin{figure*}
    \centering
\label{FigGam2}%
\end{figure*}
\renewcommand{\thefigure}{9}
\begin{figure}[!ht]
   \includegraphics[angle=270,width=\hsize]{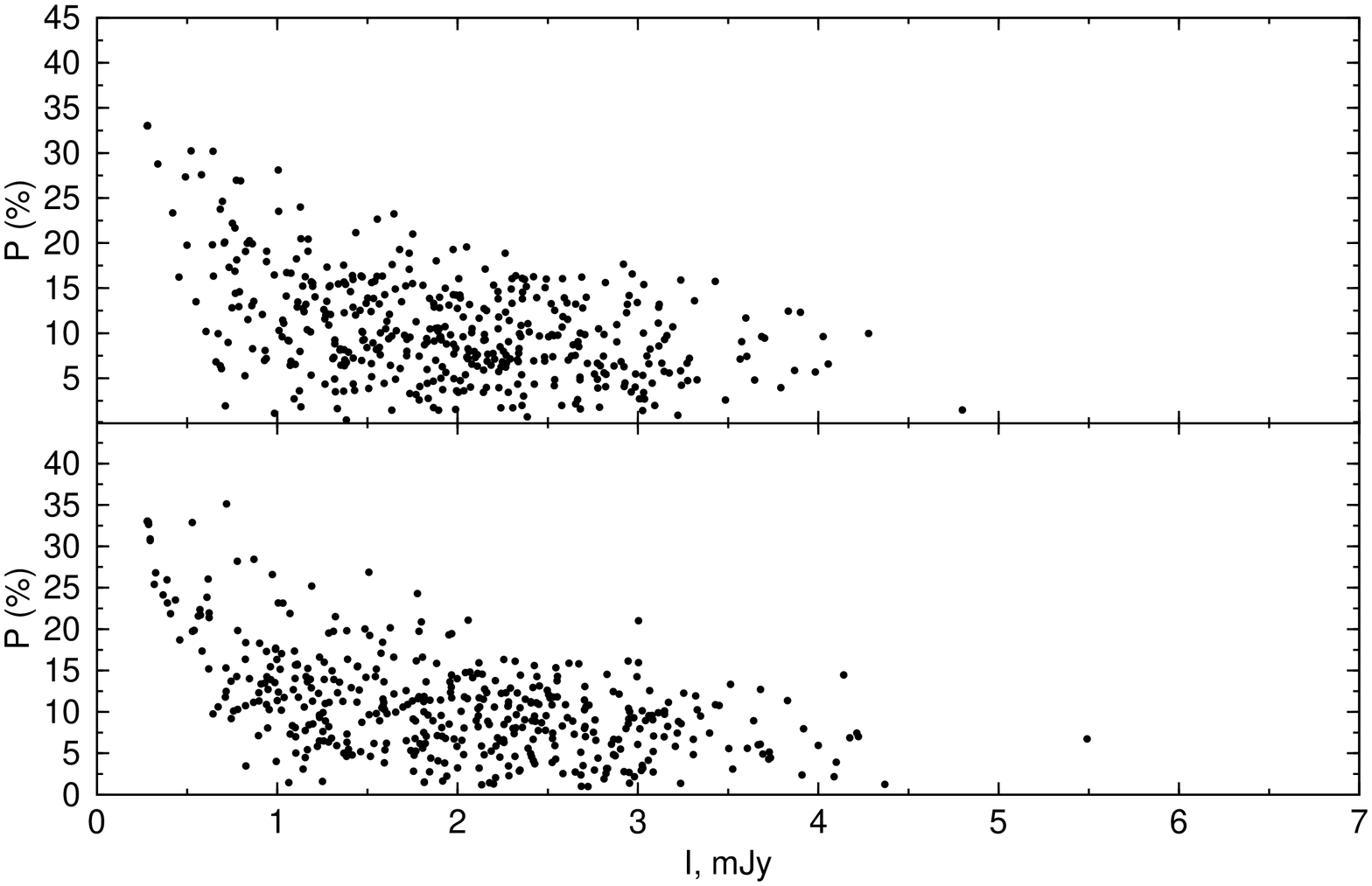}
   \caption{The dependence of the degree of polarization on flux for two different realizations.}
               \label{P_Iadd}%
   \end{figure}

\renewcommand{\thefigure}{10}
\begin{figure}[!ht]
   \includegraphics[angle=270,width=\hsize]{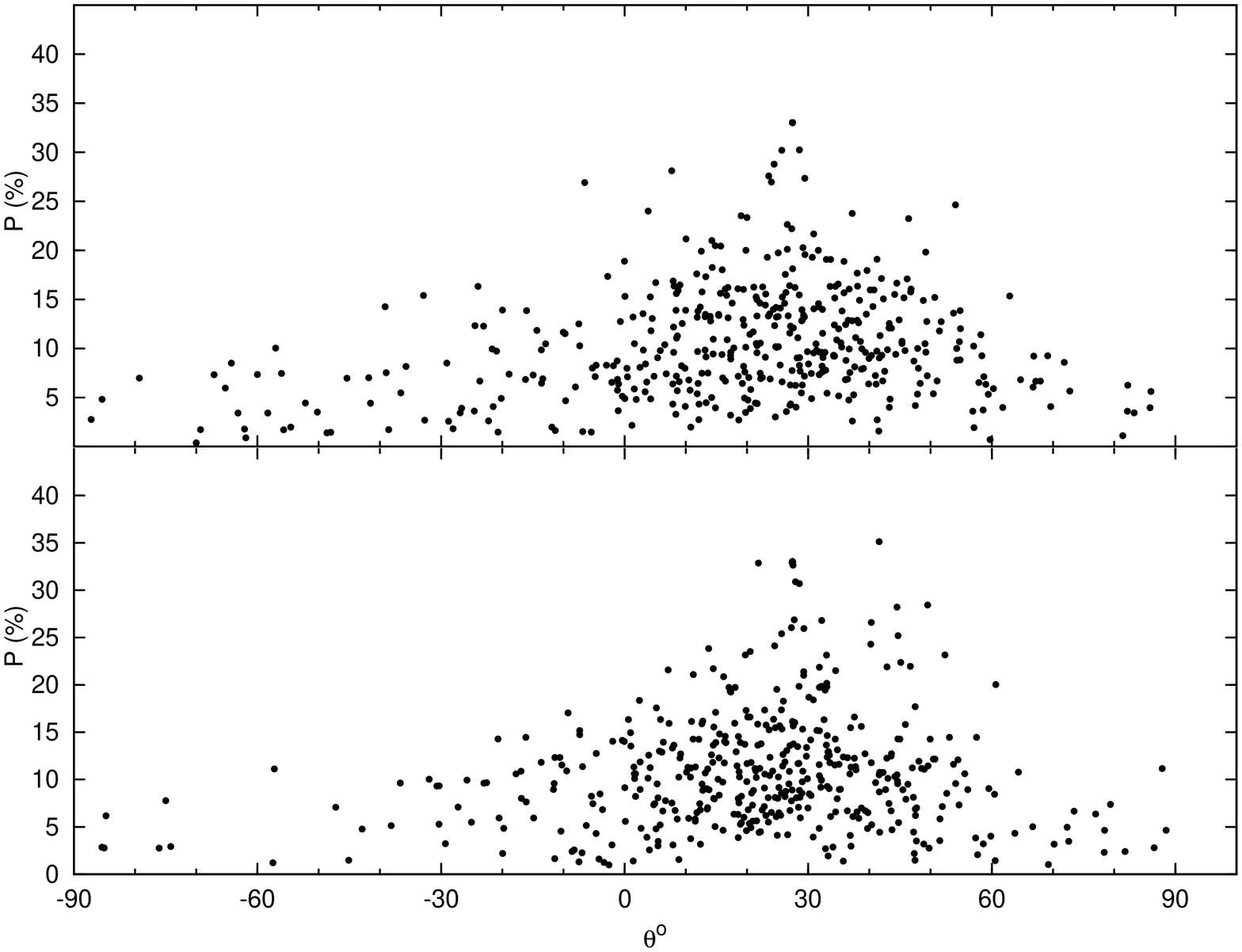}
   \caption{The dependence of the degree of polarization on its direction for two different realizations.}
   \label{P_Thadd}
\end{figure}

\begin{acknowledgements}
The work was supported by RFBR grant № 09-02-00092.
\end{acknowledgements}

\end{document}